\documentclass[12pt,preprint]{aastex}

\def\s{{\rm s}}

\def\km{{\rm\,km}}

\def\au{{\rm AU}}
\def\deg{{\rm deg}}

\def\yr{{\rm\,yr}}

\def\pomega{\tilde{\omega}}

\begin{document}

\shortauthors{Chiang}
\shorttitle{Kuiper Belt Family}

\title{A Collisional Family in the Classical Kuiper Belt}

\author{E.~I.~Chiang}

\affil{Center for Integrative Planetary Sciences\\
Astronomy Department\\
University of California at Berkeley\\
Berkeley, CA~94720, USA}

\email{echiang@astron.berkeley.edu}

\begin{abstract}
The dynamical evolution of Classical Kuiper Belt Objects (CKBOs) divides
into two parts,
according to the secular theory of test particle orbits.
The first part is a forced oscillation driven by
the planets, while the second part is a free oscillation whose
amplitude is determined by the initial orbit of the test
particle. We extract the free orbital inclinations and free orbital
eccentricities
from the osculating elements of 125 known CKBOs. The free inclinations of 32
CKBOs strongly cluster
about $2^{\circ}$ at orbital semi-major axes between 44 and 45 AU.
We propose that these objects comprise a collisional family,
the first so identified in the Kuiper Belt. Members of this family
are plausibly the fragments of an ancient parent body having
a minimum diameter of $\sim$800 km. This body was disrupted
upon colliding with a comparably sized object, and generated
ejecta having similar free inclinations.
Our candidate family is dynamically akin to a sub-family of Koronis
asteroids located at semi-major axes less than 2.91 AU; both families
exhibit a wider range in free eccentricity
than in free inclination, implying that the relative velocity
between parent and projectile prior to impact lay mostly in the
invariable plane of the solar system. We urge more discoveries
of new CKBOs to test the reality of our candidate family and
physical studies of candidate family members to probe the heretofore
unseen interior of a massive, primitive planetesimal.

\end{abstract}


\section{INTRODUCTION}
\label{intro}

According to linear secular theory, the inclination, $i$,
and longitude of ascending node, $\Omega$, of the orbit
of a test particle embedded in a planetary system
evolve with time, $t$, as

\begin{eqnarray}
p  \equiv  i \sin \Omega & = & p_{\mathit{forced}} \; + \; p_{\mathit{free}}
\nonumber \\
\label{p}
 & = & \sum_{i=1}^N A_i \sin (f_i t + \gamma_i) \; + \; i_{\mathit{free}} \sin
(Bt + \gamma_{\mathit{free}}) \\
q  \equiv  i \cos \Omega & = & q_{\mathit{forced}} \; + \; q_{\mathit{free}}
\nonumber \\
\label{q}
 & = & \sum_{i=1}^N A_i \cos (f_i t + \gamma_i) \; + \; i_{\mathit{free}} \cos
(Bt + \gamma_{\mathit{free}})
\end{eqnarray}

\noindent [see, e.g., Chapter 7 of Murray \& Dermott (1999)]. Here $f_i$
and $\gamma_i$ are functions of the masses, orbital semi-major axes,
inclinations, and nodes of the planets; $A_i$ depends on these same quantities
and the semi-major axis of the particle; and $B$ is given by the planetary
masses and orbital semi-major axes of the planets and of the particle.
The number of eigenmodes describing the forced vertical motion equals
$N$ and is usually equal to the number of planets in the system.
The ``free'' or ``proper'' inclination, $i_{\mathit{free}}$, and
$\gamma_{\mathit{free}}$
are constants specified by the initial $i$ and $\Omega$. Analogous
equations exist for the test particle's eccentricity, $e$, and longitude
of periastron, $\pomega$.

Hirayama (1918) noted that certain asteroids share similar values
of $i_{\mathit{free}}$, $e_{\mathit{free}}$, and semi-major axis, $a$. He
termed
these groups ``families,'' and christened the first three families
Koronis, Eos, and Themis using the names of their respective first-discovered
members. A family's members are thought to be fragments of a collisionally
disrupted parent body. Immediately following the disruption event,
each fragment is characterized by osculating orbital elements
$a_0 + \delta a$, $e_0 + \delta e$, and $i_0 + \delta i$, where
the subscript $0$ pertains to the parent body just prior to
disruption. Fragments for which $\delta a/a_0$, $\delta e/e_0$, $\delta i/i_0
\ll 1$ have nearly the same $i_{\mathit{free}}$ and $e_{\mathit{free}}$. A
non-zero $\delta a$
induces a change in the free precession frequency, $\delta B$.
As the orbits of fragments differentially precess in their nodes and apses,
their osculating $i$'s and $e$'s may span wide ranges, particularly if
the magnitude of the forced term is close to that of the free term.
However, the constants of integration, $i_{\mathit{free}}$ and
$e_{\mathit{free}}$,
will remain fixed barring non-secular effects (e.g., close encounters
with planets, or mean-motion resonant effects), effectively recording
the inceptive shattering event.

Classical Kuiper Belt Objects [CKBOs;
see, e.g., the reviews by Jewitt \& Luu (2000),
Farinella \& Davis (2000), and Malhotra, Duncan, \& Levison (2000)],
which we define to be those bodies having orbital semi-major axes
between 43 and 47 AU, currently do not reside within low-order mean-motion
resonances with the giant planets. Secular theory well
describes the dynamical evolution of CKBOs on low-$e$, Neptune-avoiding
orbits, modulo the effects of weak chaos
(Torbett 1989; Torbett \& Smoluchowski 1990; Duncan, Levison, \& Budd 1995).
Here we report on an elementary calculation of the free inclinations
and free eccentricities of known CKBOs. We uncover evidence for the
first collisional family in trans-Neptunian space. The candidate family is
dynamically reminiscent
of a sub-family of the Koronis population in the asteroid belt.
Section \ref{method} describes
our computational method, and section \ref{results}
contains results and discussion.

\section{METHOD}
\label{method}
Osculating elements of CKBOs were downloaded from the Minor Planet
Center (MPC; \url{http://cfa-www.harvard.edu/iau/lists/TNOs.html}).
We selected
only those objects observed at multiple apparitions and having
fitted semi-major axes $43 < a(\au ) < 47$. See Millis et al.~(2002)
for a discussion of how orbital uncertainties depend on the arclength
of astrometric observations. The resultant list contained 125 CKBOs.

For each CKBO having an observed $p_{\mathit{obs}} \equiv i_{\mathit{obs}} \sin
\Omega_{\mathit{obs}}$
and $q_{\mathit{obs}} \equiv i_{\mathit{obs}} \cos \Omega_{\mathit{obs}}$, we
compute

\begin{equation}
i_{\mathit{free}} = \sqrt{ (p_{\mathit{obs}} - p_{\mathit{forced}})^2 +
(q_{\mathit{obs}} - q_{\mathit{forced}})^2}
\end{equation}

\noindent and similarly for $e_{\mathit{free}}$. To evaluate the forced terms,
we employ the secular theory
of Brouwer \& van Woerkom (1950; hereafter BvK) for all eight
planets of the solar system, as transcribed
by Murray \& Dermott (1999).
Now the osculating elements of KBOs given by the MPC pertain to
epochs between 1994 and 2002 A.D., and are referred to the ecliptic
and equinox of J2000. Since the BvK solution defines $t = 0$
to be 1900 A.D., we take $t = 100$ yrs in equations (\ref{p})
and (\ref{q}) and ignore the small, 8-year spread in MPC epochs.
Since the BvK solution refers to the ecliptic and equinox of B1950,
we transform the MPC orbital angles ($i$, $\Omega$, $\pomega$) to the B1950
frame
using formulae provided by Montenbruck (1989, pages 18--19).
Differences between untransformed and transformed angles are
no more than 1$^{\circ}$. We note also typographical errors
in the transcription of the BvK solution by Murray \& Dermott
(1999); in their captions for Tables 7.2 and 7.3,
$e_{ij}$ and $I_{ij}$ should instead read $e_{ji}$ and $I_{ji}$,
respectively, in keeping with their notation throughout the rest
of chapter 7 (subscript $j$ denotes the planet, while subscript
$i$ denotes the eigenmode).

Our method for extracting proper elements is crude compared
with modern techniques. We neglect terms higher than degree 2
in planetary eccentricities and inclinations; see Milani \&
Kne\v{z}evi\'{c} (1994) to remedy this deficiency.
We retain only terms that are linear in the eccentricity and inclination of
the CKBO, forsaking the semi-numerical approach of Williams (1969)
and more recent, purely numerical approaches (Kne\v{z}evi\'{c}
and Milani 2001) that avoid expansions of the test particle's
orbit altogether. Thus, we cannot expect accuracy in our results for
highly inclined and eccentric CKBOs. Nonetheless, many observed
CKBOs execute nearly circular and co-planar trajectories
for which we expect the errors introduced by our analysis to be small.
And for simplicity and ease
of implementation, our procedure is probably bested only by
Hirayama's celebrated and successful (1918) analysis, which included only
a truncated Jovian secular potential [$N$ = 1 in equations
(\ref{p}) and (\ref{q})]. It is in his pioneering
spirit that we search for the first collisional family in the Kuiper Belt.

\section{RESULTS AND DISCUSSION}
\label{results}

\begin{figure}
\epsscale{0.8}
\plotone{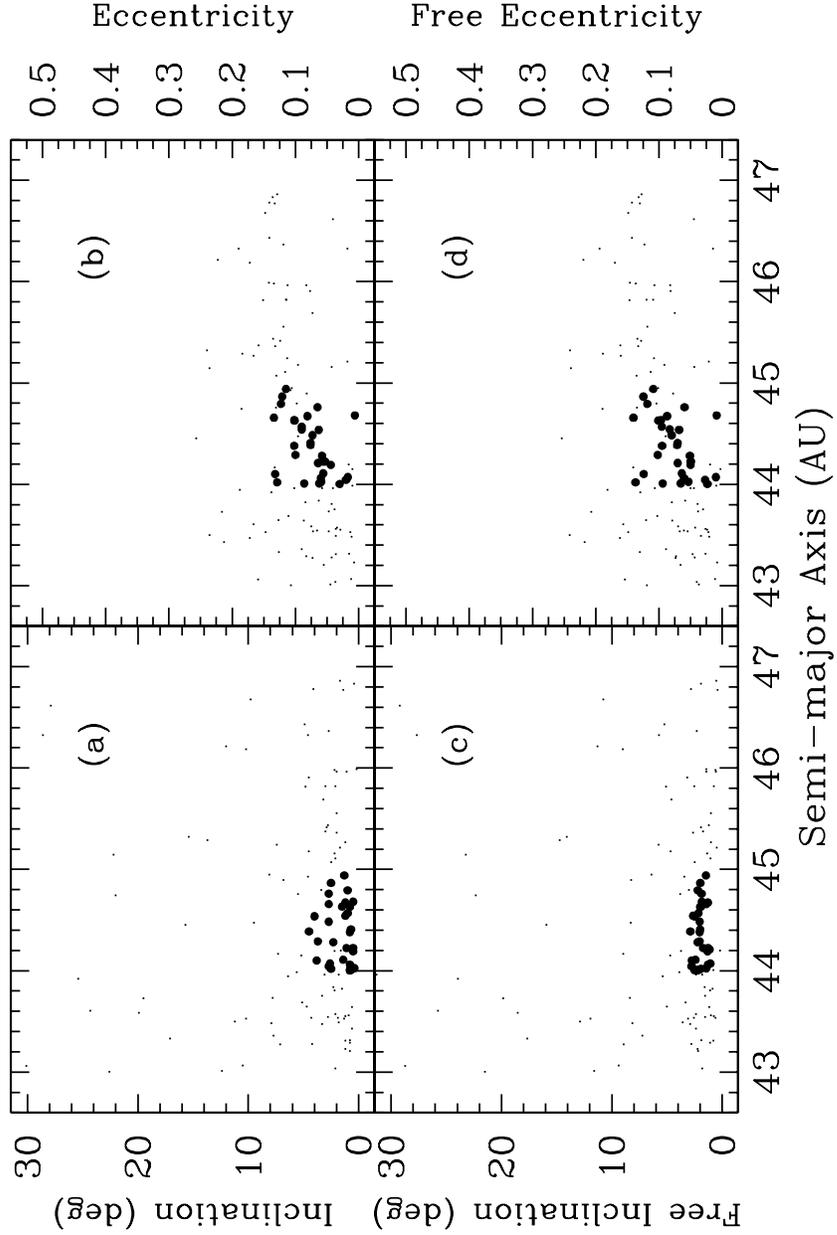}
\caption{(ab) Observed (osculating) inclinations and eccentricities
of 125 multi-opposition CKBOs, as supplied by the Minor Planet Center
on March 25, 2002. (cd) Free inclinations and eccentricities
of these objects, as computed from observed values,
using the secular theory of
Brouwer \& van Woerkom (1950) to subtract off forced contributions.
Emboldened circles represent the 32 objects having $44 < a(\au) < 45$
and $1 < i_{\mathit{free}} (\deg) < 3$. The free inclinations
cluster more strongly than the observed inclinations, pointing
to a common, likely collisional origin for these objects.
The spread in $e_{\mathit{free}}$ of our candidate family
is larger than in $i_{\mathit{free}}$,
similar to behavior exhibited by a sub-family of the Koronis asteroids.
\label{free}}
\end{figure}

Figure \ref{free} displays the inclinations
and eccentricities, observed and free, of CKBOs against their semi-major axes.
While the observed elements betray no obvious clumping, this
is not true for the free inclinations.
We consider the population of KBOs having $44 \lesssim a(\au) \lesssim 45$
and $1 \lesssim i_{\mathit{free}} (\deg) \lesssim 3$ to be a candidate
collisional family. Table 1 provides the designations, semi-major axes, free
elements, and $H$-magnitude inferred diameters
of our 32 candidate family members.

The free inclinations of our candidate family cluster
about $2^{\circ}$, similar to the inclination
of the invariable plane with respect to the ecliptic,
$i_{\mathit{invar}}$. This coincidence might lead one
to suspect that these bodies represent primordial ones
that coagulated in the invariable plane,
rather than fragments generated by catastrophic disruption.
In this alternative scenario, these bodies would
appear in $p$-$q$ space at the point representing the
inclination and node of the invariable plane referred to the ecliptic.
Would all such bodies share the same $i_{\mathit{free}}$,
without recourse to collisional genesis?
There are two ways by which they might do so,
but both paths are prohibitive.
The first way is if $i_{\mathit{forced}} \ll i_{\mathit{free}}$
(see Figure 7.3 in Murray \& Dermott), so that
$i_{\mathit{free}} \approx i_{\mathit{invar}}$.
The former condition fails to be satisfied; for
every one of our candidate members,
$i_{\mathit{forced}} \approx i_{\mathit{free}}$.
The second option demands that all bodies appear
in $p$-$q$ space over a timescale much shorter than
the local forced precession timescale of
$\sim$$2\pi/f_8 \approx 2 \times 10^5\yr$. In this case, the
forced inclination vectors of bodies appearing in
$p$-$q$ space would be nearly identical since there would
be insufficient time for the forced vectors
to rotate. There would be differences in the amplitudes
of the forced inclination vectors from body to body, but these would be
small since $\Delta a / a$ is small for our candidate
family and the bodies are far from secular resonances.
Consequently, the free inclination vectors
of all bodies appearing at a single point in $p$-$q$ space
would be nearly the same.
However, for all members of our candidate family to have formed well within
a time interval of $2 \times 10^5 \yr$ seems unreasonable.
Current planetesimal coagulation calculations indicate KBO
growth times of order $10^7\yr$ (Kenyon 2002, and references therein).
By contrast, a collisional disruption event is virtually instantaneous;
we proceed under the hypothesis of a collisional origin for
these objects.

Our candidate family boundaries were chosen based on
visual inspection of Figure \ref{free}c
and the remarkable narrowing of the distribution
of inclinations between panels \ref{free}a
and \ref{free}c exhibited by our candidate family.
A Kolmogorov-Smirnov (K-S) test comparing the distributions
of observed and free inclinations
for our candidate family yields a 0.1\% probability
that the two distributions are drawn from the
same underlying distribution. We consider this probability
sufficiently low that we believe the narrowing
of the inclination distribution to be significant.
No other subset of KBOs in Figure \ref{free}
evinces the same degree of sharpening
when we transform from observed to free elements;
we compute corresponding K-S probabilities of $\sim$20\% for
other subsets of KBOs.

Comparing the distribution
of free inclinations of our candidate family
to that of 22 KBOs having $43 \lesssim a(\au) \lesssim 44$
and $1 \lesssim i_{\mathit{free}} (\deg) \lesssim 4$
reveals no significant difference; the K-S test
yields a 38\% probability that the two distributions
are drawn from the same population. A comparison
between our candidate family and the 12 KBOs
having $45 \lesssim a(\au) \lesssim 46$
and $1 \lesssim i_{\mathit{free}} (\deg) \lesssim 3$
gives a corresponding probability of 13\%, again too high
to claim a significant difference. Thus, our
family, if real, might extend over the full gamut of
semi-major axes from 43 to 46 AU.
Subjectivity in the choice of family boundaries is
a problem that afflicts even the relatively mature field
of asteroidal dynamics, and clearly more objects
would be desirable. We defer more sophisticated analyses
of the significance of the clumping in $i_{free}$-$a$ space
to future study. For the present, we reserve our attention to the
32 KBOs between $a = 44 \au$ and 45 AU that exhibit
the sharpest concentration of free inclinations.

\placefigure{fig1}

The free eccentricities of our candidate family are more broadly
distributed than their free inclinations. This characteristic
is reminiscent of the Koronis asteroid sub-family located
at semi-major axes $2.84 \lesssim a (\au) \lesssim 2.91$. This
sub-family exhibits
an absolute width of $\sim$0.025 in $e_{\mathit{free}}$ and an absolute width
of $\sim$0.005 in $\sin i_{\mathit{free}}$ (Bottke et al.~2001).\footnote{The
Koronis family divides into two groups (``sub-families''), one
located at $2.84 \lesssim a (\au) \lesssim 2.91$ and a second
located at $2.92 \lesssim a (\au) \lesssim 2.95$. The free
eccentricities in the latter sub-family are markedly higher
than those in the former, reflecting excitation via passage
through a secular resonance located at 2.92 AU (Bottke et al.~2001).
The free eccentricities and free inclinations of the former
sub-family are thought to be of collisional origin,
and we employ this sub-family as the analogue of the Kuiper Belt family
proposed here. Within the former Koronis sub-family, the spread in free
eccentricities is still larger than the spread in free inclinations,
and appears to demand an anisotropic distribution of ejecta velocities.}
The corresponding widths for our candidate Kuiper Belt family are 0.15 and
0.03, respectively. That these widths are each $\sim$6 times
greater than their Koronis sub-family counterparts probably reflects
the greater ease with which orbital trajectories are altered in the more
distant Kuiper Belt; orbital velocities are $\sim$4 times
lower in the CKB than in the asteroid belt. The larger
variation in $e_{\mathit{free}}$ compared to that in $i_{\mathit{free}}$
exhibited by both families presumably reflects a collision in which
fragment ejecta velocities were greater in the (invariable)
plane of the solar system than out of the plane. This, in turn,
probably implicates a collision between two ancient bodies
having greater relative velocity in the plane than out of the plane.

\placefigure{fig2}
\begin{figure}
\epsscale{0.8}
\plotone{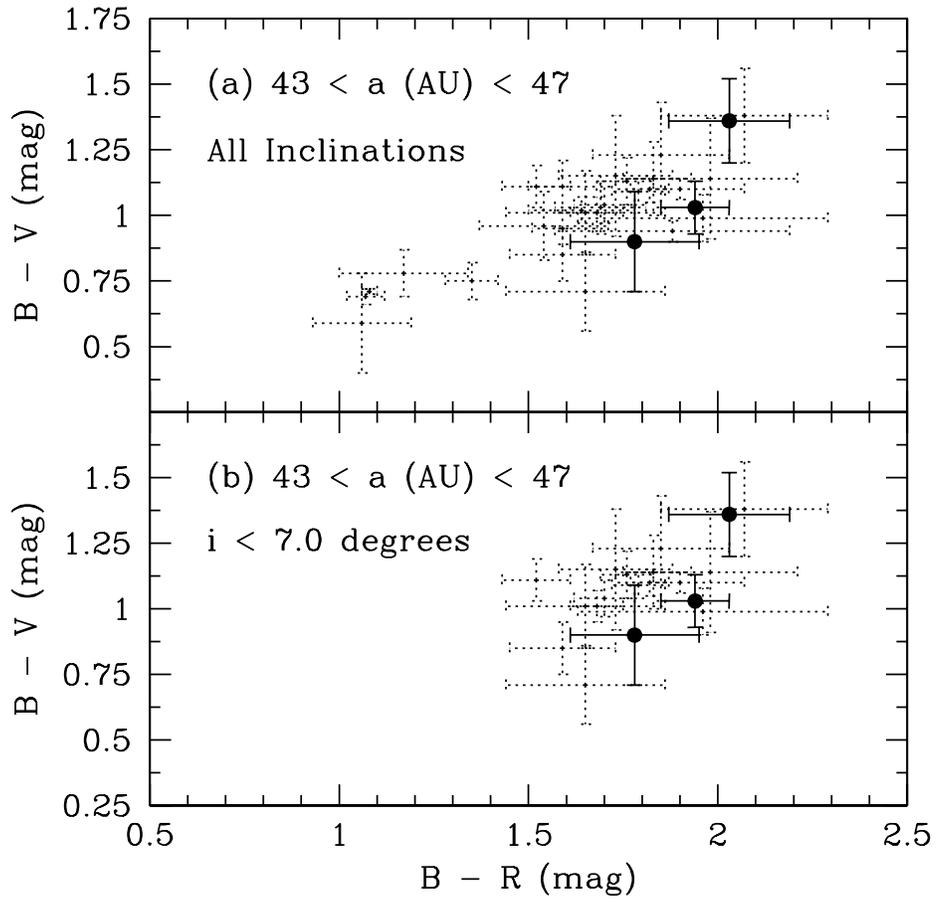}
\caption{(a) Optical colors of CKBOs. Only 3 of the 32 candidate
family members in Table 1 have had their colors measured; they are
symbolized by solid circles.
(b) Same as (a), except that only colors of CKBOs having
$i_{\mathit{obs}} < 7^{\circ}$ are plotted.
\label{color}}
\end{figure}

The mass of the projectile, $M_{\mathit{proj}}$, that shattered the
parent body could not have been much smaller than that of the parent,
$M_{\mathit{par}}$. For the kinetic energy of the collision (measured in
the frame moving at the center-of-mass velocity) to exceed the
gravitational binding energy of the parent body,
$\min (M_{\mathit{proj}}/M_{\mathit{par}}) \sim
(v_{\mathit{esc,par}}/v_{\mathit{rel}})^2$, where
$v_{\mathit{esc,par}}$ is the escape velocity from the parent body and
$v_{\mathit{rel}}$ is the relative velocity between projectile and target.
Using the fragment sizes in Table 1 that are derived
from an assumed albedo $A = 0.04$, we estimate
the minimum diameter of the putative parent body to
be $\sim$$800 \sqrt{0.04/A} \, \km$, for which
$\min (v_{\mathit{esc,par}}) \sim 0.4 \sqrt{0.04/A} \, \km \; \s^{-1}$. The
relative velocity
must be a fraction, $f$, of the local orbital velocity:
$v_{\mathit{rel}} \sim 1 \, (f/0.2) \, \km \; \s^{-1}$. Thus,
$\min (M_{\mathit{proj}}/M_{\mathit{par}}) \sim 0.2 \, (0.2/f)^2 \, (0.04/A)$.

Surface reflectance spectra of members of a dynamical family
can lend supporting evidence for a common physical ancestry. The Koronis
sub-family
exhibits relatively homogeneous S-type optical colors
that set it apart from the more variegated field population
(e.g., Gradie, Chapman, \& Williams 1979). The parent body
for the Koronis family is thought to be $\sim$$120\km$ in diameter (Bottke
et al.~2001), too small to have undergone significant internal differentiation.
More massive parent bodies may have differentiated; collisional
fragments from such bodies may display greater variation
in their spectral properties. Thus, finding a large spread
in optical colors among members of a given candidate family does not
necessarily eliminate the candidate from the running;
widely different colors and/or spectra may instead offer a glimpse
into the shattered, chemically zoned interior of a minor planet.

Unfortunately, color photometry exists for only 3 of our 32 candidate family
members
(1999 CM119, 1999 HU11, and 1999 CO153) in Table 1.
Their B-V and B-R colors are plotted
in Figure \ref{color}, along with the available colors of all other
CKBOs [we employ data from Trujillo \& Brown (2002), selecting only those
objects having $43 < a(\au) < 47$]. The tendency for the 3 candidate
family members to be redder than those of the general Classical
population (Figure \ref{color}a) may imply homogeneity
of the interior of the putative parent body. Alternatively, it may reflect
whatever unknown process is responsible for generating the color-inclination
correlation reported by Trujillo \& Brown (2002). In an attempt
to separate out this effect, we plot only data for those CKBOs having
$i_{\mathit{obs}} < 7^{\circ}$ in
Figure \ref{color}b. There remains
a tendency for our KBO family candidates to skirt
only the lower envelope of the space spanned by all
low-$i$ CKBOs, though clearly there are too few points
and the uncertainties in individual points are too large
to claim significant segregation. We urge physical studies of
the candidate family members listed in Table 1 to further
explore the possibility that we are indeed viewing the
heretofore unseen interior of a massive, primitive planetesimal.

\acknowledgements
I am grateful to Marc W. Buie and Amy B. Jordan for inspiring this calculation
during an observing run plagued by fog and hail. I thank
also an anonymous referee for a thoughtful and thought-provoking
report that significantly improved this paper. This work
was supported in part by a Junior Faculty Research Grant
awarded by the University of California at Berkeley.

\newpage
\begin{deluxetable}{lcccc}
\tablewidth{0pt}
\tablecaption{Candidate Kuiper Belt Family Members}
\tablehead{
\colhead{MPC Designation} &
\colhead{a (AU)} & \colhead{$i_{\mathit{free}}$ (deg)} &
\colhead{$e_{\mathit{free}}$} & \colhead{Diameter\tablenotemark{a} (km)}
}
\startdata
2001 ES24 &	44.537 &	2.49 &	0.068 &	175 \\
2001 DB106 &	44.101 &	2.77 &	0.124 &	332 \\
2000 YV1	&	44.629 &	1.99 &	0.101 &	253 \\
2000 PN30 &	44.632 &	1.95 &	0.097 &	160 \\
2000 PY29 &	44.061 &	1.27 &	0.060 &	231 \\
2000 GY146 &	44.044 &	2.78 &	0.027 &	175 \\
2000 GX146 &	44.408 &	2.01 &	0.071 &	183 \\
2000 GV146 &	44.290 &	2.04 &	0.102 &	201 \\
2000 FR53 &	44.866 &	1.98 &	0.124 &	192 \\
2000 FG8	&	44.224 &	1.72 &	0.049 &	183 \\
2000 FC8	&	44.009 &	2.13 &	0.066 &	167 \\
2000 FA8	&	44.003 &	2.30 &	0.023 &	210 \\
2000 CH105 &	44.542 &	2.64 &	0.083 &	303 \\
2000 CF105 &	44.191 &	1.33 &	0.050 &	277 \\
2000 CE105 &	44.210 &	1.15 &	0.070 &	290 \\
2000 CL104 &	44.673 &	1.31 &	0.087 &	381 \\
1999 RA216 &	44.380 &	2.05 &	0.095 &	241 \\
1999 RE215 &	44.940 &	1.47 &	0.109 &	317 \\
1999 RC215 &	44.108 &	2.43 &	0.064 &	277 \\
1999 OF4	&	44.760 &	1.87 &	0.060 &	277 \\
1999 OZ3	&	44.019 &	1.93 &	0.137 &	220 \\
1999 HU11\tablenotemark{b} &	44.025 &	1.43 &	0.053 &	303 \\
1999 HS11 &	44.071 &	1.09 &	0.001 & 332 \\
1999 DH8	&	44.387 &	2.89 &	0.071 &	122 \\
1999 CU153 &	44.484 &	2.05 &	0.080 &	241 \\
1999 CS153 &	44.794 &	2.21 &	0.119 &	160 \\
1999 CO153\tablenotemark{b}&	44.008 &	2.53 &	0.094 &	220 \\
1999 CN119 &	44.568 &	2.16 &	0.096 &	167 \\
1999 CM119\tablenotemark{b} &	44.657 &	1.44 &	0.140 &	175 \\
1999 CC119 &	44.680 &	1.85 &	0.009 & 253 \\
1998 HM151 &	44.225 &	1.29 &	0.050 &	167 \\
1995 DC2	&	44.281 &	2.23 &	0.051 &	241 \\
\enddata
\tablenotetext{a}{Based on an assumed albedo $A = 0.04$ and an absolute
$H$-magnitude supplied by the Minor Planet Center.}
\tablenotetext{b}{B-V and B-R colors are available for these objects.}
\end{deluxetable}

\end{document}